# Three years of SPHERE: the latest view of the morphology and evolution of protoplanetary discs

Antonio Garufi[1,2]
Myriam Benisty[3]
Tomas Stolker[4]
Henning Avenhaus[2]
Jos de Boer[5]
Adriana Pohl[6]
Sascha P. Quanz[2]
Carsten Dominik[4]
Christian Ginski[5]
Christian Thalmann[2]
Roy van Boekel[6]
Anthony Boccaletti[7]
Thomas Henning[6]
Markus Janson[6]
Graeme Salter[8]
Hans Martin Schmid[2]
Elena Sissa[9]
Maud Langlois[8]
Jean-Luc Beuzit[3]
Gaël Chauvin[3]
David Mouillet[3]
Jean-Charles Augereau[3]
Andreas Bazzon[2]
Beth Biller[6]
Mickael Bonnefoy[3]
Esther Buenzli[2]
Anthony Cheetham[10]
Sebastian Daemgen[2]
Silvano Desidera[9]
Natalia Engler[2]
Markus Feldt[6]
Julien Girard[3]
Raffaele Gratton[9]
Janis Hagelberg[3]
Christoph Keller[5]
Miriam Keppler[6]
Matthew Kenworthy[5]
Quentin Kral[7]
Bruno Lopez[11]
Anne-Lise Maire[6]
François Menard[3]
Dino Mesa[9]
Sergio Messina[9]
Michael R. Meyer[2]
Julien Milli[3]
Michiel Min[4]
André Muller[6]
Johan Olofsson[6]
Nicole Pawellek[6]
Christophe Pinte[3]
Judit Szulagyi[2]
Arthur Vigan[8]
Zahed Wahhaj[8]
Rens Waters[4]
Alice Zurlo[8]

[1] Universidad Autónoma de Madrid, Departamento de Física Teórica, Madrid, Spain
[2] Institute for Astronomy, ETH Zurich, Switzerland
[3] Université Grenoble Alpes, Institut de Planétologie et d'Astrophysique de Grenoble, France
[4] Anton Pannekoek Institute for Astronomy, University of Amsterdam, the Netherlands
[5] Sterrewacht Leiden, the Netherlands
[6] Max Planck Institute for Astronomy, Heidelberg, Germany
[7] LESIA, Observatoire de Paris, PSL Research Université, CNRS, Université Paris Diderot, Sorbonne Paris Cité, UPMC Paris 6, Sorbonne Universités, France
[8] Aix Marseille Université, CNRS, Laboratoire d'Astrophysique de Marseille, UMR 7326, France
[9] INAF – Osservatorio Astronomico di Padova, Italy
[10] Geneva Observatory, University of Geneva, Versoix, Switzerland
[11] Laboratoire Lagrange, Université Côte d'Azur, Observatoire de la Côte d'Azur, CNRS, Nice, France

Spatially resolving the immediate surroundings of young stars is a key challenge for the planet formation community. SPHERE on the VLT represents an important step forward by increasing the opportunities offered by optical or near-infrared imaging instruments to image protoplanetary discs. The Guaranteed Time Observation Disc team has concentrated much of its efforts on polarimetric differential imaging, a technique that enables the efficient removal of stellar light and thus facilitates the detection of light scattered by the disc within a few au from the central star. These images reveal intriguing complex disc structures and diverse morphological features that are possibly caused by ongoing planet formation in the disc. An overview of the recent advances enabled by SPHERE is presented.

The large number of exoplanets discovered with various techniques (for example, transit, radial velocity, imaging) indicates that planet formation around young stars is very common. The architecture of the observed planetary systems is surprisingly varied and typically does not resemble that of our Solar System. For example, planets with masses intermediate between our terrestrial and gaseous planets are very frequent. Also, orbital distances smaller than that of our innermost planet are well populated, whereas only a few giant planets have also been found in orbits that are much larger than Neptune's (for example, HR 8799b, HIP 65426b).

This variety likely points to a great diversity in the initial conditions of planet formation throughout all evolutionary stages, from dust grain growth to planet-disc interactions and possibly planet migration. However, the mechanisms of planet formation are only partly understood. Different scenarios have been proposed that are not necessarily mutually exclusive and that act over a range of timeframes. A better characterisation of the physical and chemical properties of protoplanetary discs is therefore crucial in understanding planet formation.

The Spectro-Polarimetric High-contrast Exoplanet REsearch instrument (SPHERE), described in Beuzit et al. (2008) began operating on the Very Large Telescope (VLT) in May 2014. Since February 2015, the team responsible for the Guaranteed Time Observation (GTO) programme (principal investigator: Jean-Luc Beuzit) has dedicated substantial effort and observing time to the characterisation of circumstellar discs, with approximately ten complete nights of telescope time dedicated to debris discs (for example, Lagrange et al., 2016) and a similar amount of time dedicated to protoplanetary discs.

## Protoplanetary discs with SPHERE

While optical and near-infrared observations of stars at a distance of a few hundred parsecs may naturally guarantee excellent spatial resolution, imaging the circumstellar environments at these wavelengths suffers from the low contrast of the circumstellar emission compared to the stellar flux. The most efficient way to mitigate this effect is to use a form of differential imaging. In particular, extended structures like circumstellar discs at tens of au from the star are best imaged with

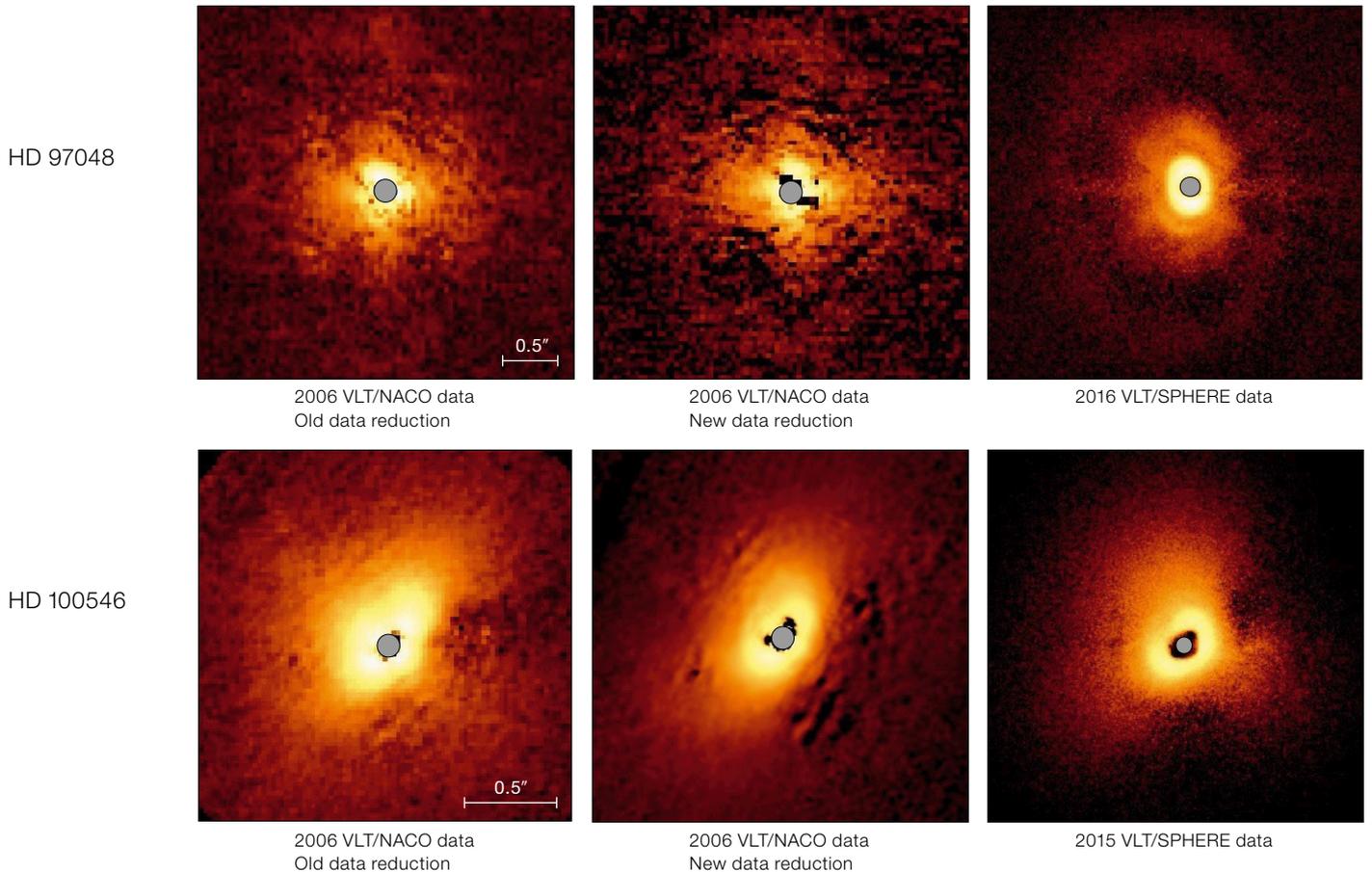

HD 97048

2006 VLT/NACO data
Old data reduction

2006 VLT/NACO data
New data reduction

2016 VLT/SPHERE data

HD 100546

2006 VLT/NACO data
Old data reduction

2006 VLT/NACO data
New data reduction

2015 VLT/SPHERE data

Figure 1. The improving data quality with the new reduction procedures and the impact of SPHERE are illustrated for the two disc systems HD 97048 and HD 100546. The HD 97048 image from NACO old reduction is from Quanz et al. (2012), and the NACO new reduction and SPHERE data are from Ginski et al. (2016). The HD 10056 images are from, left to right, Quanz et al. (2011), Avenhaus et al. (2014) and Garufi et al. (2016).

polarimetric differential imaging (PDI), which exploits the different nature of the stellar (mainly unpolarised) light and scattered (strongly polarised) light from the disc surface. The (quasi-)simultaneous observation of orthogonal polarisation states enables the separation of these two light components, which are then combined to obtain the full set of linear Stokes parameters, as well as the azimuthal component $Q_\phi$ of the polarised light (see Schmid et al., 2006).

Until five years ago, the NAOS-CONICA (NACO) instrument on the VLT was the only ESO facility offering PDI. Some pioneering work gave us a hint of the morphological diversity of discs (see, for example, the review by Quanz et al., 2011). However, there was no systematic approach to the observation of discs using PDI, with only a handful of objects observed at the time. The performance of PDI has significantly improved over the last five years thanks to: (i) an optimised data reduction, which corrects for the instrumental polarisation and employs the azimuthal $Q_\phi$ parameter to reduce the noise floor in the image (see Avenhaus et al., 2014a); and (ii) the advent of SPHERE. Figure 1 illustrates the effects of these advances.

SPHERE was mainly designed to provide imaging and spectroscopic characterisation of exoplanets. However, SPHERE's extreme adaptive optics (XAO) module SAXO (Fusco et al., 2006) and the high-precision polarimetry offered by the subsystems — the Zurich IMaging POLarimeter ZIMPOL (Thalmann et al., 2008) and the InfraRed Dual-band Imager and Spectrograph IRDIS (Dohlen et al., 2008) — make SPHERE among the best instruments to perform PDI of circumstellar discs. This combination equips it with an unprecedented combination of excellent sensitivity (more than four orders of magnitude in contrast) and angular resolution (~ 3 au in the visible and ~ 7 au in the near-infrared for sources at 150 pc). Furthermore, it is the only instrument mounted on an AO-assisted 8-metre telescope to perform PDI in the visible.

Imagery of protoplanetary discs

A glimpse of the morphological variety of protoplanetary discs emerging from the GTO programme can be seen in Figure 2. These ten sources represent only a small fraction of the entire sample of discs imaged so far with PDI. Nonetheless, all the morphological elements now routinely discovered in all protoplanetary discs can be appreciated here. Our census, initially biased towards more luminous stars, is now more uniform across all stellar types, with discs around T Tauri stars like LkCa 15 (Thalmann et al., 2016) being as well characterised as the most imaged Herbig Ae/Be systems, like HD 100546 (Garufi et al., 2016).

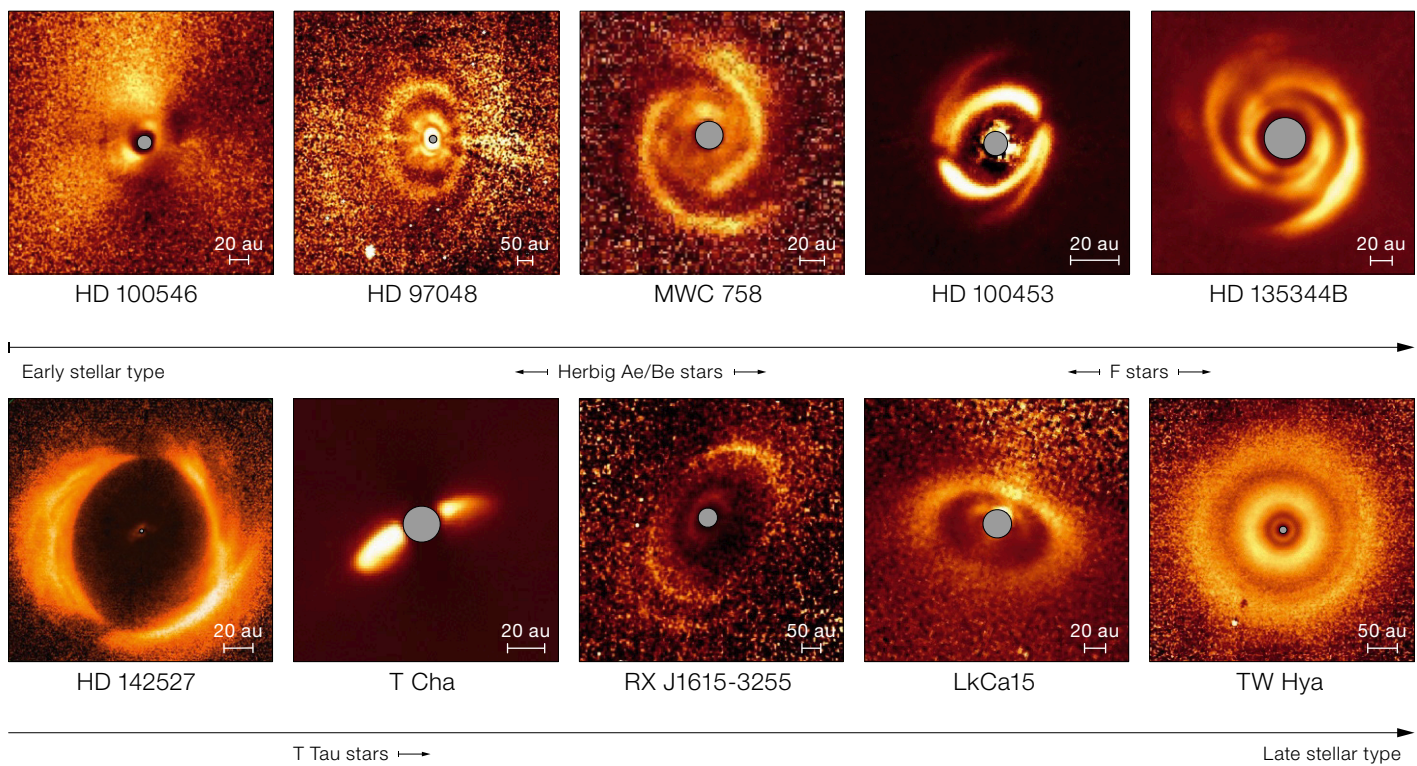

Figure 2. Collection of images of protoplanetary discs observed in PDI with SPHERE. References to the images are: HD 100546, Garufi et al. (2016); HD 97048, Ginski et al. (2016); MWC 758, Benisty et al. (2015); HD 100453, Benisty et al. (2017); HD 135344B, Stolker et al. (2106); HD 145527, Avenhaus et al. (2017); T Cha, Pohl et al. (2017); RX J1615-3255, de Boer et al. (2016); LkCa 15, Thalmann et al. (2016); TW Hya, van Boekel et al. (2017).

The clearest finding from an inspection of the available sample is that all discs show morphological features. In the majority of sources, either concentric rings (HD 97048, Ginski et al., 2016; TW Hya, van Boekel et al., 2017) or spiral arms (MWC 758, Benisty et al., 2015; HD 135344B, Stolker et al., 2016a) are revealed. Some discs, predominantly those with spirals, also show radially extended dips that can be interpreted as shadows cast by a misaligned disc at a few au from the central star (Benisty et al., 2017; Avenhaus et al., 2017).

The interpretation of features from inclined discs is less immediate because of the degeneracy between scattering phase function, disc geometry and illumination effects in these sources. This analysis is nonetheless pivotal to constrain the composition of dust grains (Stolker et al., 2016b; Pohl et al., 2017) and the geometry of the disc surface (de Boer et al., 2016), both of which are necessary to understand planet formation.

### Comparison with ALMA images

PDI images are sensitive to micron-sized dust grains at the disc surface. These grains are very well coupled to the gas under typical disc conditions. On the other hand, images at (sub-)millimetre wavelengths trace larger grains within the disc. Comparing SPHERE and Atacama Large Millimeter/Submillimeter Array (ALMA) images with comparable angular resolution can potentially reveal the different morphologies of different disc components. In fact, many disc processes (for example, grain growth or dust filtration) are expected to differentiate the distribution of gas and large grains throughout disc evolution, leaving their imprint on the disc structure. This is illustrated in Figure 3 for two prototypical examples from PDI, one showing concentric rings (TW Hya) and another showing spiral arms (HD 135344B).

Similarly to the PDI data, the ALMA images of TW Hya show a number of rings and gaps (Andrews et al., 2016). Van Boekel et al. (2017) performed a detailed comparison of the radial profiles of these two datasets, highlighting both similarities and profound differences. The entire detectable signal from ALMA is located within the second bright ring from PDI at approximately 60 au. The two main millimetre dips seen in both the image and the radial profile in Figure 3 have analogous dips in PDI at 25 au and at 40 au. Similar considerations apply to some rings. In general, large-scale structures have a stronger contrast in SPHERE data, whereas narrow features appear more profound with ALMA. There is no general consensus regarding the origin of these rings, with both planet-disc interactions and dust accumulation in correspondence with ice lines being the most promising explanations.

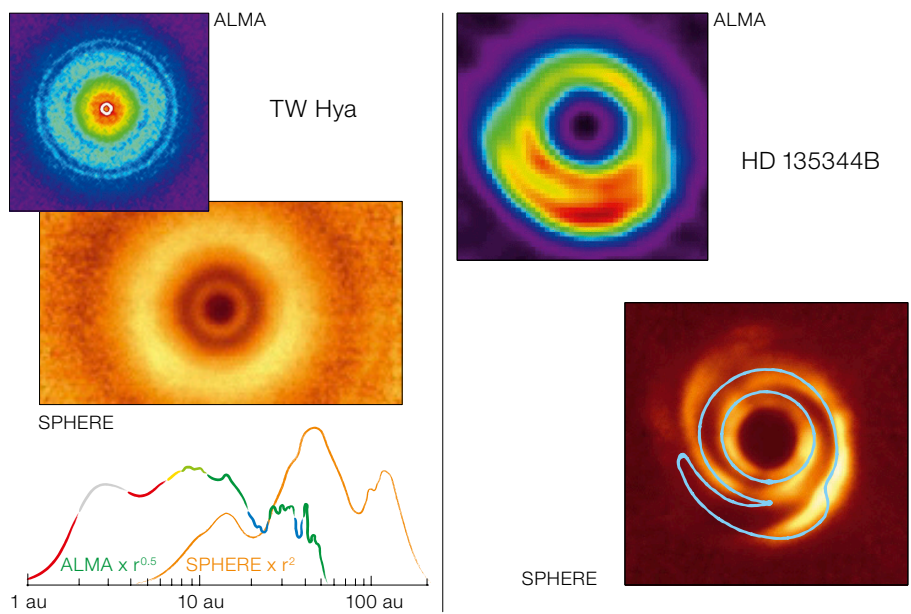

Figure 3. Comparison of ALMA and SPHERE images for two prototypical discs shown at the same spatial scale. For TW Hya, the ALMA image is from Andrews et al. (2016) and the SPHERE image and radial profiles from van Boekel et al. (2017). For HD 135344B, the ALMA image is from van der Marel et al. (2016) and the SPHERE image is from Stolker et al. (2016).

On the other hand, the spiral arms visible in the PDI image of HD 135344B are not detected by ALMA (van der Marel et al., 2016) even though the lower angular resolution of these images still leaves this scenario open. The best available image of this source shows an inner ring at approximately 45 au and an asymmetric feature to the south running parallel to the ring. The inner ring lies at a larger distance than the inner rim in scattered light at 25 au. The southern feature does not match the location of the two spirals but rather seems to follow the trail of the western one, or alternatively to run parallel to the eastern one but at a larger orbital radius. These radial and azimuthal displacements between small and large grains may be speculatively related to the formation history of these features, with a vortex-like structure to the south being generated by an inner dust trap at smaller radii, which is in turn related to the possible presence of planets within the inner cavity (van der Marel et al., 2016).

## Disc cavity or no disc cavity

Another fundamental aspect revealed by both near-infrared and millimetre images is the high occurrence of large (> 10 au) central cavities. In many cases, these cavities appear smaller in PDI than in the millimetre images (for example, HD 135344B). In some other cases, they are not detected in PDI down to the instrument coronagraph, even though this is up to four times smaller in size than the millimetre cavity (for example, MWC 758). A systematic study of these differences is fundamental in the context of planet-disc interactions, since the most probable scenario for the formation of these cavities is the dynamical action of orbiting companions.

The high occurrence of cavities in discs surrounding Herbig stars can be appreciated from Figure 4. It can be seen that the long-standing observational dichotomy found by Meeus et al. (2001) between sources with large far-infrared excess (Group I) and small far-infrared excess (Group II) can be understood in terms of the presence or absence of a large disc cavity. From the sample of Herbig stars with available PDI images (covering the vast majority of sources within 200 pc), it is clear that discs with large cavities represent more than half of the total. If we exclude low-mass discs where outer stellar companions at 100s of au are likely truncating the disc from outside (bottom-left of the chart) this fraction increases to ~ 75 %.

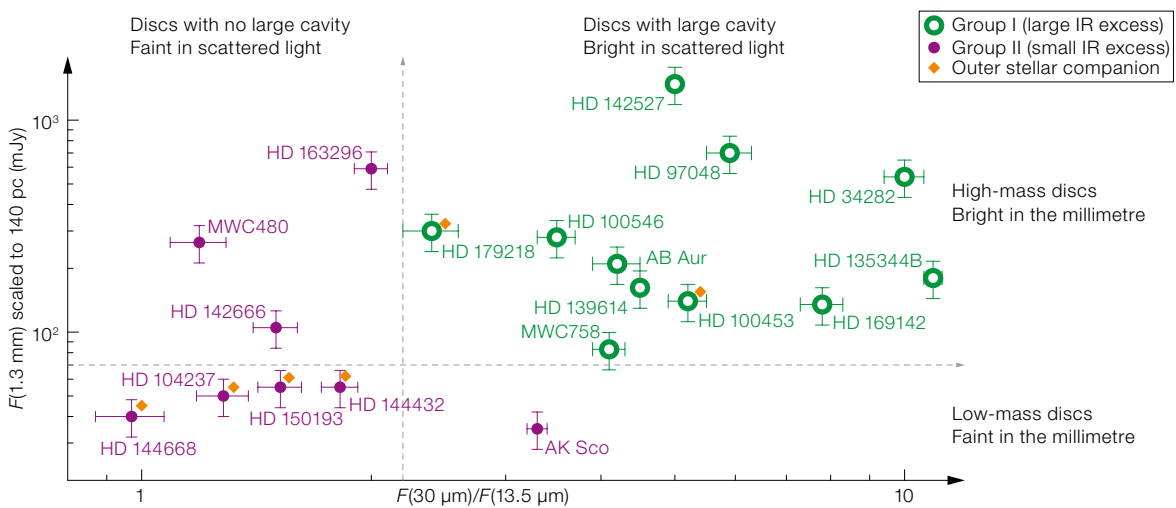

Figure 4. Millimetre to mid-infrared chart (adapted from Garufi et al., 2017). The photometric ratio of 30 μm to the 13.5 μm ratio on the x-axis is a proxy for the amount of scattered light traced by SPHERE. The y-axis indicates the millimetre brightness scaled to the same distance. The brown diamonds denote sources with outer stellar companions.

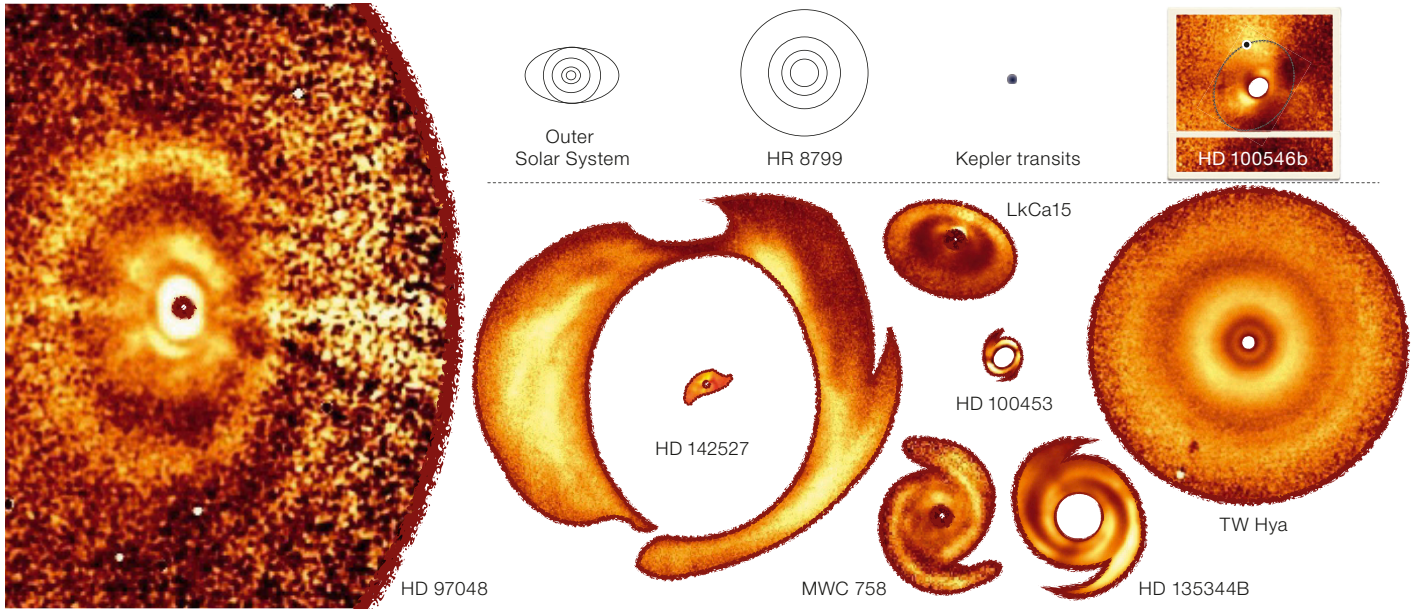

Figure 5. Illustrative protoplanetary discs and planetary systems all shown at the same physical scale.

Figure 4 also shows that discs without a large cavity (Group II) are fainter in scattered light. The photometric ratio of the x-axis is in fact a good proxy for the amount of signal detected in PDI (Garufi et al., 2017) while the millimetre flux along the y-axis gives an approximate measurement of the dust mass. Figure 4 shows that Group II sources can be divided into two sub-categories according to their millimetre flux, likely explaining the elusiveness of Group II sources in scattered light. Millimetre flux observations reveal both discs as massive as Group I, which are likely shadowed by their own inner rim at the (sub-)au scale, and those with low mass and an outer companion, which are probably smaller than (or comparable with) the size of the coronagraph employed for the PDI observations and are thus faint.

Both the presence and the size of the disc cavity, as well as the disc radial extent, may be intimately related to the morphology of the resulting planetary systems. However, a taxonomic approach to the resolved information of the discs

Figure 6. Stellar mass – age diagram for the Herbig stars within 200 pc, compared with the pre-main sequence tracks by Siess et al. (2000), for a given spectral type and varying luminosity.

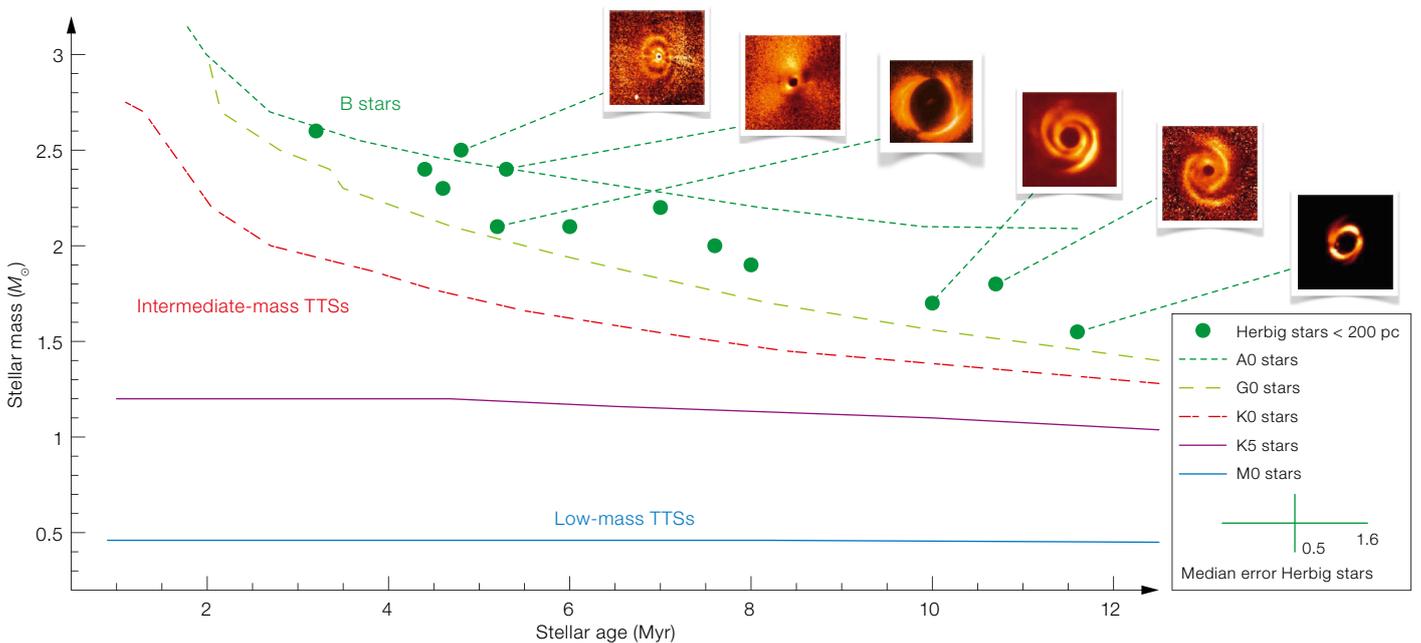

and the architecture of planetary systems has yet to be attempted.

### Link with planetary systems

Unfortunately, the intermediate stages of planet formation cannot be studied observationally because of the lack of detectable emission from metre- to kilometre-sized bodies. Thus, the most obvious approach to establishing a link between protoplanetary discs and planetary systems is the direct comparison of their physical scale, as shown in Figure 5. The planetary orbits in systems like that of HR 8799, or our own Solar System, lie at scales comparable to the locations of some disc features like rings, cavities and spirals.

However, the vast majority of the currently known exoplanets have been discovered by the Kepler space mission with the transit method, which is strongly biased towards planets within a few au of the central star. To date, very few gas giant planets have been imaged beyond 30 au (for example, the SPHERE SHINE survey; Chauvin et al., 2017). In other words, the routinely discovered disc features do not have an obvious counterpart in the exoplanet distribution. This can be explained by one or more of the following considerations:
(a) Planet migration due to disc forces is a very efficient and rapid process;
(b) A large population of as yet unseen low-mass planets exists at large orbits;
(c) The discs that we observe are exceptional, as with HR 8799;
(d) Commonly observed disc features are not due to the interaction with planets.

In any case, the solution to this apparent incongruity is related to the mechanisms of planet formation. Disc fragmentation due to instability is believed to be a possible mechanism in the early evolution of discs (less than 1 Myr old), whereas the stars we observe with discs are much older (see below). Accretion of gas onto planetary embryos as massive as a few Earth masses appears more likely at this stage. In this scenario, disc gaps and cavities may be sculpted by (accreting?) giant planets and/or facilitate the formation of kilometre-sized planetesimals from the intense growth of millimetre-sized particles (through a dust trap).

The most promising observational approach to these questions is the detection and characterisation of forming planets still embedded in the disc, and possibly of their circumplanetary disc. To date, very few such candidates have been found. Among these, HD 100546b is one of the best studied, even though its impact on the disc morphology remains somehow elusive (see for example, Quanz et al., 2013; Garufi et al., 2016).

### Evolution of discs (and of our view of it)

Owing to large uncertainties in stellar ages, there have not been conclusive results on the evolution of disc features with time. The new Gaia distances to Herbig stars enable us to refine the stellar properties, as depicted in Figure 6. In this diagram the stellar masses and ages of the nearby (< 200 pc), and therefore most studied, Herbig stars (i.e., with spectral types between F6 and B9), are compared to pre-main sequence tracks of later (cooler) spectral-type stars.

Bearing in mind both the small number of objects and the uncertainty in pre-main sequence models, it is clear that discs showing symmetric double-spiral arms are found around older (i.e., less luminous) stars, like HD 135344B and MWC 758. On the other hand, very extended discs with multiple and more complex structures are found predominantly around younger (more luminous) stars, like HD 97048 and HD 100546.

Another interesting hint from the diagram in Figure 6 is that all well-studied Herbig stars within 200 pc are older than 3 Myr. To find younger stars still more massive than the Sun, we should look for: (i) early B stars, (ii) particularly luminous A and F stars, or (iii) intermediate-mass T Tauri stars (IMTTS; namely T Tauri stars with spectral types between G0 and K5). Unfortunately, no star from the first two categories has been found within 200 pc. This means that to study the younger counterparts of the well-known discs around Herbig stars, we should image the surroundings of IMTTS, especially in the more luminous systems. From the diagram, it is also clear how a larger range of stellar masses is covered with a certain spectral type interval when moving to later spectral types.

All of these considerations motivate us to extend our past successful study of Herbig stars to T Tauri stars. Fortunately, SPHERE performs much better than expected with fainter stars. We have already imaged the circumstellar environment of more than 40 T Tauri stars, both in GTO and in open-time programmes. The next step for the consortium will be the comparison of these newly imaged disc features around T Tauri stars with those known from older and/or more massive systems. Our goal is to compare the physical conditions for planet formation to carry out a complete census of stellar ages and masses.